\documentclass[useAMS,usenatbib]{mnras}
\usepackage{graphicx,url}
\usepackage{amsmath}
\usepackage{color}

\def\msun{$M_{\odot}$}

\def\xstar{{\tt XSTAR}}
\def\xillver{{\tt xillver}}
\def\relxill{{\tt relxill}}
\def\relxilllp{{\tt relxilllp}}
\def\relline{{\tt relline}}

\def\msun{$M_{\odot}$}

\begin{document}

\title[High-Density Effects on X-ray Reflection]{The Effects of High Density on 
the X-ray Spectrum Reflected from Accretion Disks around Black Holes}
\author[Garc\'ia et. al.]
{Javier~A.~Garc\'ia$^1$\thanks{E-mail: javier@head.cfa.harvard.edu},
        Andrew~C.~Fabian$^2$,
        Timothy~R.~Kallman$^4$,
        Thomas~Dauser$^3$,
        \newauthor Michael~L.~Parker$^2$,
        Jeffrey~E.~McClintock$^1$,
        James~F.~Steiner$^5$,
        J\"orn Wilms$^3$\\
$^1$Harvard-Smithsonian Center for Astrophysics,
  60 Garden St., Cambridge, MA 02138 USA\\
$^2$Institute of Astronomy, Madingley Road, Cambridge, CB3 0HA, UK\\
$^3$Remeis Observatory \& ECAP, Universit\"at
  Erlangen-N\"urnberg, Sternwartstr.~7, 96049 Bamberg, Germany\\
$^4$NASA Goddard Space Flight Center, Code 662, Greenbelt, MD 20771, USA\\
$^5$MIT Kavli Institute for Astrophysics and Space Research, MIT, 70 Vassar Street, Cambridge, MA 02139, USA\\
}

\maketitle

%

\begin{abstract}

Current models of the spectrum of X-rays reflected from accretion
disks around  black holes and other compact objects are commonly
calculated assuming that the density of the disk atmosphere is constant
within several Thomson depths from the irradiated surface. An important
simplifying assumption of these models is that the ionization structure
of the gas is completely specified by a single, fixed value of the
ionization parameter $\xi$, which is the ratio of the incident flux to
the gas density.  The density is typically fixed at
$n_e=10^{15}$~cm$^{-3}$. Motivated by observations, we consider higher
densities in the calculation of the reflected spectrum. We show by computing model spectra for
$n_e\ga10^{17}$~cm$^{-3}$ that high-density effects significantly modify
reflection spectra. The main effect is to boost the thermal continuum at
energies $\la 2$~keV. We discuss the implications of these results
for interpreting observations of both AGN and black hole binaries. We
also discuss the limitations of our models imposed by the quality of the
atomic data currently available.

\end{abstract}


%
%
%
%
\section{Introduction}\label{sec:intro}

X-ray reflection spectroscopy is arguably the most effective means
currently available for probing the effects of strong gravity near the
event horizon of an accreting black hole.  The reflection spectrum is
produced from the reprocessing of high-energy coronal photons in an
optically-thick accretion disk. The result is a rich spectrum of
radiative recombination continua, absorption edges and fluorescent
lines, most notably the Fe K complex in the 6--8~keV energy range
\citep{ros93,roz02,ros93,gar10}.  This reflected radiation leaves the
disk carrying a wealth of information on the physical composition and
condition of the matter in the strong field near a black hole. The Fe K
emission lines (and other fluorescent lines) are broadened and shaped by
Doppler effects, light bending and gravitational redshift
\citep{fab00,rey03,dov04,dau12}.

Reflection models are the {\it sine qua non} of the Fe-line method of
measuring black hole spin, which is enormously important because of its
dominant role in measuring the spins of supermassive black holes in
active galactic nuclei (AGN)
\citep[e.g.,][]{pat11,wal13,rey14,par14}. Importantly, this method is equally
applicable to measuring the spins of stellar-mass black holes \citep[i.e.,
black hole binaries, BHB;][]{mcc14,gou14,ste14,mil15,che15}. X-ray reflection is also
observed in other astrophysical sources such as neutron stars \citep{cak10},
cataclysmic variables \citep{muk15}, and ultra-compact X-ray binaries
\citep{mad14}.

For the past three decades, models of X-ray reflection, which are
complex, have undergone continual improvement \citep[see][for a
review]{fab10}. Currently, the most advanced model is \xillver, which
relative to earlier models incorporates a more complete atomic database
and an improved solution to the problem of radiative transfer
\citep{gar10,gar13a}. The reflection model \xillver\ has been linked with
the relativistic convolution code \relline\ \citep{dau10,dau13} to
provide the complete relativistic reflection package \relxill\
\citep{gar14a}, which is presently the state-of-the-art in modeling
ionized reflection in strong gravity. As the wealth of observational
data grows in both quantity and quality, so must the theoretical models.

A common simplifying assumption of reflection models is that the density
of the disk atmosphere is constant, although hydrostatic atmosphere
models have been explored \citep[e.g.,][]{nay01,bal01,roz08}. The
assumption of constant density significantly simplifies the problem and
so far has proven to provide a good representation of the data. A second
simplification for constant density calculations is that a specific
model can be characterized by a single value of the ionization
parameter, which is proportional to the ratio of the illuminating flux
to the gas density. In photoionization calculations, it is standard
practice to assume that, for all values of flux and density under
consideration, the ionization structure of the gas is solely determined
by this one parameter. That is, it is simply the ratio of flux to
density that determines the ionization structure of the atmosphere and
properties of the reflection spectrum. 

%
%
\begin{figure}
\centering
\includegraphics[scale=0.5,angle=0,trim={2cm 0.5cm 0 0}]{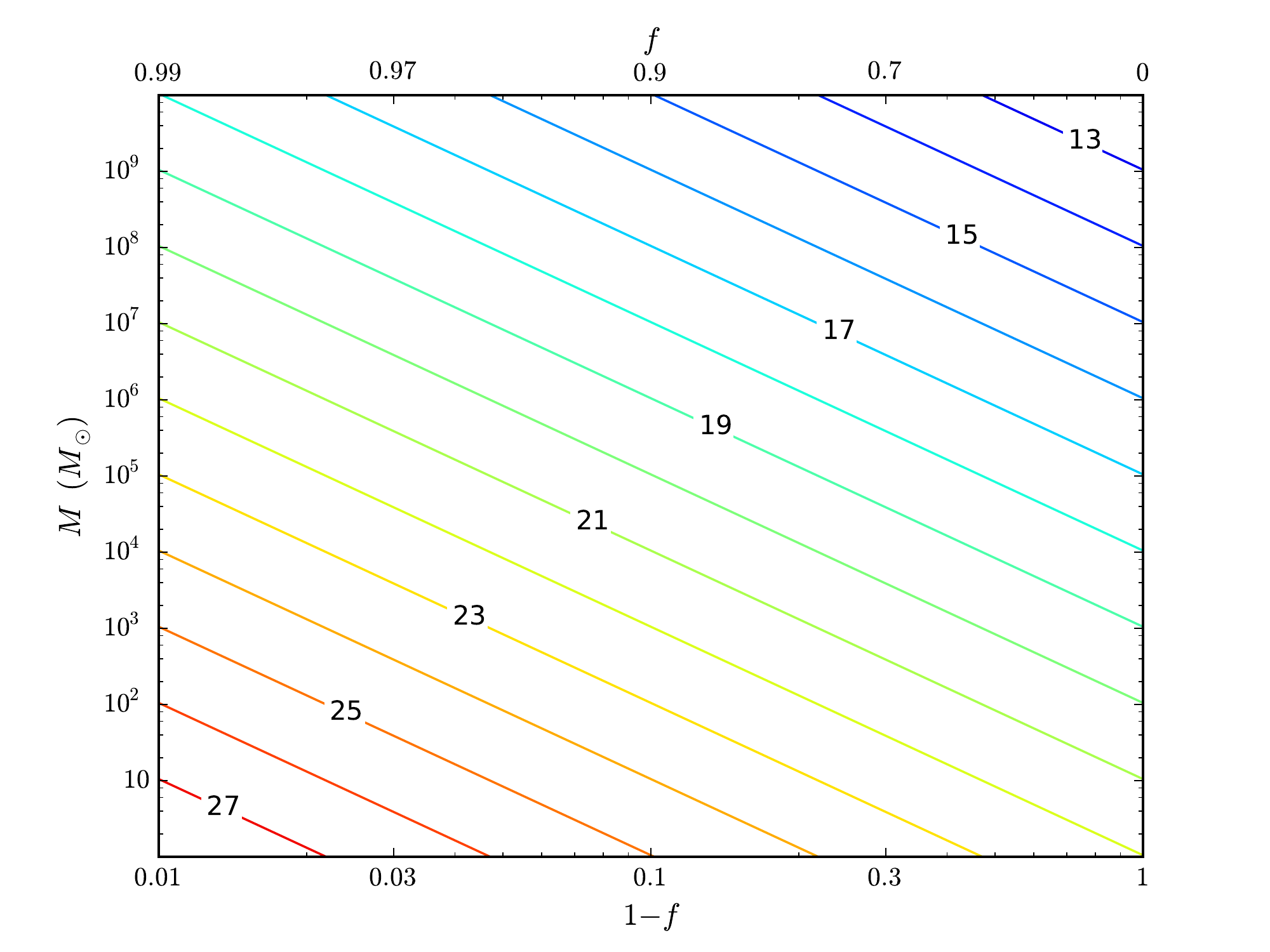}
\caption{
Black hole mass versus coronal power based on the radiation pressure
solution from \citet{sve94}. Lines show constant $n_e\dot m^2$ for different
values of the density, as indicated, and assuming $r=10$ and $\alpha=0.1$. For
a 10 solar mass black hole at the minimum $f$, the density is still near
$10^{22}$~cm$^{-3}$.
}
\label{fig:density}
\end{figure}

The effects of high density on the reflection spectrum are most
important for powerful coronae and for black holes of low mass. Some
high-density models ($n_{\rm e} = 10^{20}$~cm$^{-3}$) for black hole
binaries were computed by \cite{ros07}, and for neutron stars by 
\cite{bal04}. An equation relating the
density of a standard $\alpha$-disk at high accretion rates, for which
the inner region is radiation-pressure dominated, is given by
\cite{sve94}:
\begin{equation}
n_e=\frac{1}{\sigma_{\rm T} R_{\rm S}} \frac{256\sqrt{2}}{27}\alpha^{-1}r^{3/2} 
\dot m^{-2} [1-(3/r)]^{-1} (1-f)^{-3},
\end{equation}
where $f$ is the fraction of the total power released into the corona;
$\sigma_{\rm T}=6.64\times10^{-25}$~cm$^2$ is the Thomson cross section;
$R_{\rm S}=2GM/c^2$ is the Schwarzschild radius, $r=R/R_{\rm S}$; and
$m=\dot Mc^2L_{\rm Edd}$ is the dimensionless accretion rate expressed
in terms of the Eddington luminosity $L_{\rm Edd}\sim
10^{38}$~erg~s$^{-1}$. Figure~\ref{fig:density} shows in the $M$ vs. $f$
plane lines of constant $n_e\dot m^2$ for $r=10$ and $\alpha=0.1$. Note
that the pressure-dominated radius disappears when $f$ approaches unity
and when $\dot m<0.1$, i.e., below 1\% of the Eddington limit \citep[see
Fig.~2 of][]{sve94}.

Most current reflection models adopt a moderate value of density,
typically $n_{\rm e} = 10^{15}$~cm$^{-3}$, which is appropriate for low
values of the coronal fraction $f$, high mass accretion rates, and black
holes with $M>10^7$~\msun.  Meanwhile, for lower accretion rates and
low-mass black holes, current models should be adequate because, as we
shall demonstrate, the major effects of higher densities on the
reflection spectrum occur below 3~keV.

In this paper, using the \xillver\ code and the approximation of
constant density, we explore the effects on reflection spectra of
varying the gas density over the range $n_{\rm e} =
10^{15}-10^{19}$~cm$^{-3}$. We show that the temperature of the
atmosphere increases markedly at $n_e\ga10^{17}$~cm$^{-3}$. The main
effect on the spectrum is a significant increase in the flux at low
energies ($\la 2$~keV).

This paper is organized as follows. In Section~\ref{sec:calc} we
describe in detail the effects of high density on the thermal and
ionization balance, and hence on models of the reflection spectrum. In
Section~\ref{sec:dis} we discuss the implications of these effects for
modeling AGN and BHB spectral data, as well as the limitations of the
models themselves that are imposed by the quality of the atomic data
currently available.

%
%
\section{X-ray Reflection at High-Density}\label{sec:calc}

The calculations of the reprocessing of X-rays in illuminated accretion disks
are carried out using our reflection code \xillver.  This approach has been
extensively described in previous papers \citep{gar10,gar13a,gar14a}, therefore
here we will only provide a short overview.  \xillver\ solves the radiative
transfer in a plane-parallel (slab) geometry, using the Feautrier method as
described in \cite{mih78}. The boundary condition at the surface of the disk,
which is placed at a Thomson depth $\tau_\mathrm{T}=10^{-4}$, specifies the
incident radiation field. A second inner boundary condition specifies the
incident radiation field at the bottom of the slab, which in our calculations
is set at $\tau_\mathrm{T}=10$. The calculation of the ionization and thermal
balance is done implementing the photoionization routines from the \xstar\ code
\citep{kal01}, which incorporates the most complete atomic database for
modeling synthetic photoionized X-ray spectra. The microphysics captured by
\xillver\ is much more detailed than for any earlier code, principally because
we treat in detail the K-shell atomic properties of many astrophysically
relevant ions \citep[e.g.][]{kal04,gar05,gar09}.

An important simplification commonly adopted in reflection modeling is that
the number density $n_e$ in the illuminated atmosphere (which sits on top of a
much more optically thick, geometrically thin accretion disk), is constant.
This simplification is motivated by the need for large grid of models that can
be easily applied to real observational data. Moreover, this has proven to be sufficient to
accurately describe the reflected spectrum. Similar calculations have been
performed assuming that the atmosphere is in hydrostatic equilibrium
\citep[e.g.,][]{nay01,bal01}. However, while such calculations are much more
complex than those for constant density, it is still unclear whether the
hydrostatic approximation provides a solution significantly closer to reality
than a constant density calculation.  In this paper we do not seek to discuss
the trade off between different approximations. For the reminder we will focus on
exploring the effects of high-density in X-ray illuminated atmospheres with
constant density.

A useful quantity to describe constant density models is the ionization
parameter:
\begin{equation}\label{eq:xi}
\xi = \frac{4\pi F_\mathrm{x}}{n_e},
\end{equation}
where $F_\mathrm{x}$ is the ionizing flux in the 1--1000~Ry band. This quantity
is typically taken as an indicator of the degree of ionization in the material.
This is because it is a measure of the ratio of the photoionization rate (which
is proportional to $F_\mathrm{x}\times n_e$) to the recombination rate (which
is proportional to $n_e^2$). Thus, in photoionization modeling of constant
density plasmas it is normally assumed the $\xi$ completely characterizes each
model, regardless of the actual values of density or flux. From this, it is
clear that the assumption that both density and flux can be characterized by a
single quantity $\xi$ is valid as long as photoionization and recombination are
the dominant processes controlling the thermal and ionization balance of the
gas. However, as we show next, this assumption is invalid at sufficiently
high densities, as other physical processes become important.

%
\subsection{Temperature Solutions}\label{sec:temp}

To explore the effects of high-density on the reflected spectra, we have
carried out calculations with \xillver\ for various values of the
density while keeping the ionization parameter constant. For all these models
the illumination is described as a power-law spectrum with slope $\Gamma=2.3$,
with a sharp low-energy cutoff at 0.1~keV and a high-energy exponential
roll-over at 300~keV.  The abundances of all elements are set to their solar
values. These parameters are typical for AGN or BHB in the soft-state.
The density is varied in the $n_e=10^{15}-10^{19}$~cm$^{-3}$ range, and
the net incident flux is also varied such that the ionization
parameter is always $\xi=50$~erg~cm~s$^{-1}$.

%
%
\begin{figure}
\centering
\includegraphics[scale=0.6,angle=0,trim={0.5cm 0 0 0}]{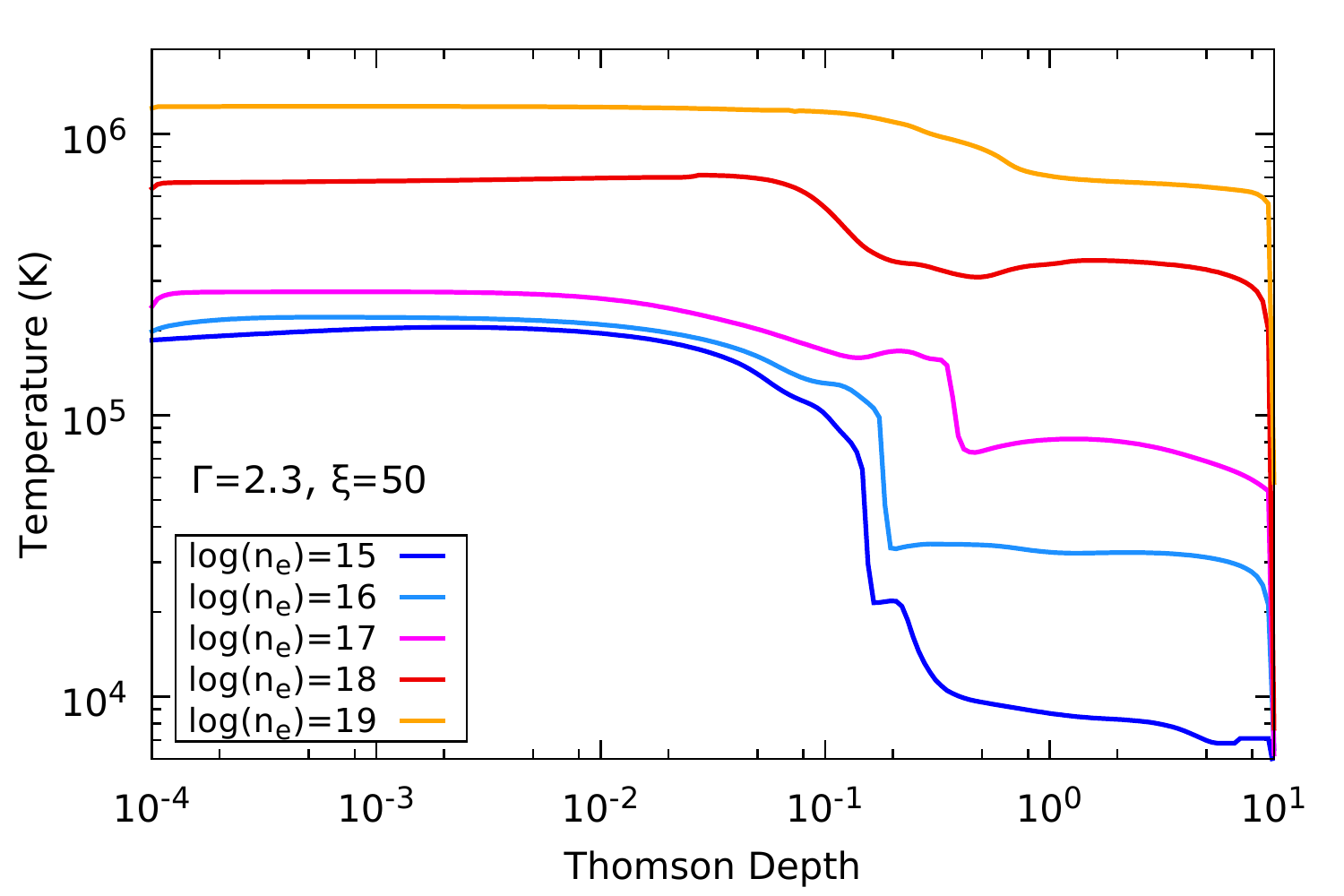}
\caption{
Temperature profiles in the vertical direction of the illuminated atmosphere.
Each curve corresponds to a different value of density, as indicated, while the
ionization parameter is held constant at $\xi=50$~erg~cm~s$^{-1}$. Other model
parameters are: $\Gamma=2.3$, $E_\mathrm{cut}=300$~keV, and $A_\mathrm{Fe}=1$.
}
\label{fig:HDtemper}
\end{figure}

Figure~\ref{fig:HDtemper} shows the temperature profiles in the vertical
direction of the illuminated slab resulting from models with five different
densities (as indicated). The overall profiles follow closely the expected
behavior in a constant density photoionized slab: the temperature is higher at
the illuminated surface ($\tau_\mathrm{T}=10^{-4}$) where the gas is more
ionized. At larger depths, as enough photons are removed from the ionizing
continuum, the gas recombines and the temperature decreases rapidly. This
two-zone profile has been reported in several previous reflection calculations
\citep[e.g.][]{ros05,gar10,gar13a}.

The temperature tends to increase with the increase in density. For densities
in the $15 < \log(n_e) < 17$ range, the hot zone of the slab shows modest
variation, while in the deeper zones the temperature rises with density. For
higher densities, the temperature increase can be observed everywhere in the
atmosphere. At each point, the temperature is determined by the balance of
Compton heating and cooling, free-free heating and cooling, photoionization
heating, and radiative recombination cooling.

Because of its quadratic dependence on the density, free-free
(Bremsstrahlung) heating-cooling is expected to be the dominant process leading
temperature changes. Another major effect of high densities is the
enhancement of the collisional de-excitation, which in turn suppresses
radiative cooling, thus raising the temperature. Additionally, three-body
recombination becomes important at $\log(n_e)\sim19$, which lowers the
ionization state. Lower ionization translates into larger photoionization
heating.

\begin{figure*}
\centering
\includegraphics[scale=0.55,angle=0,trim={0.5cm 0 0 0}]{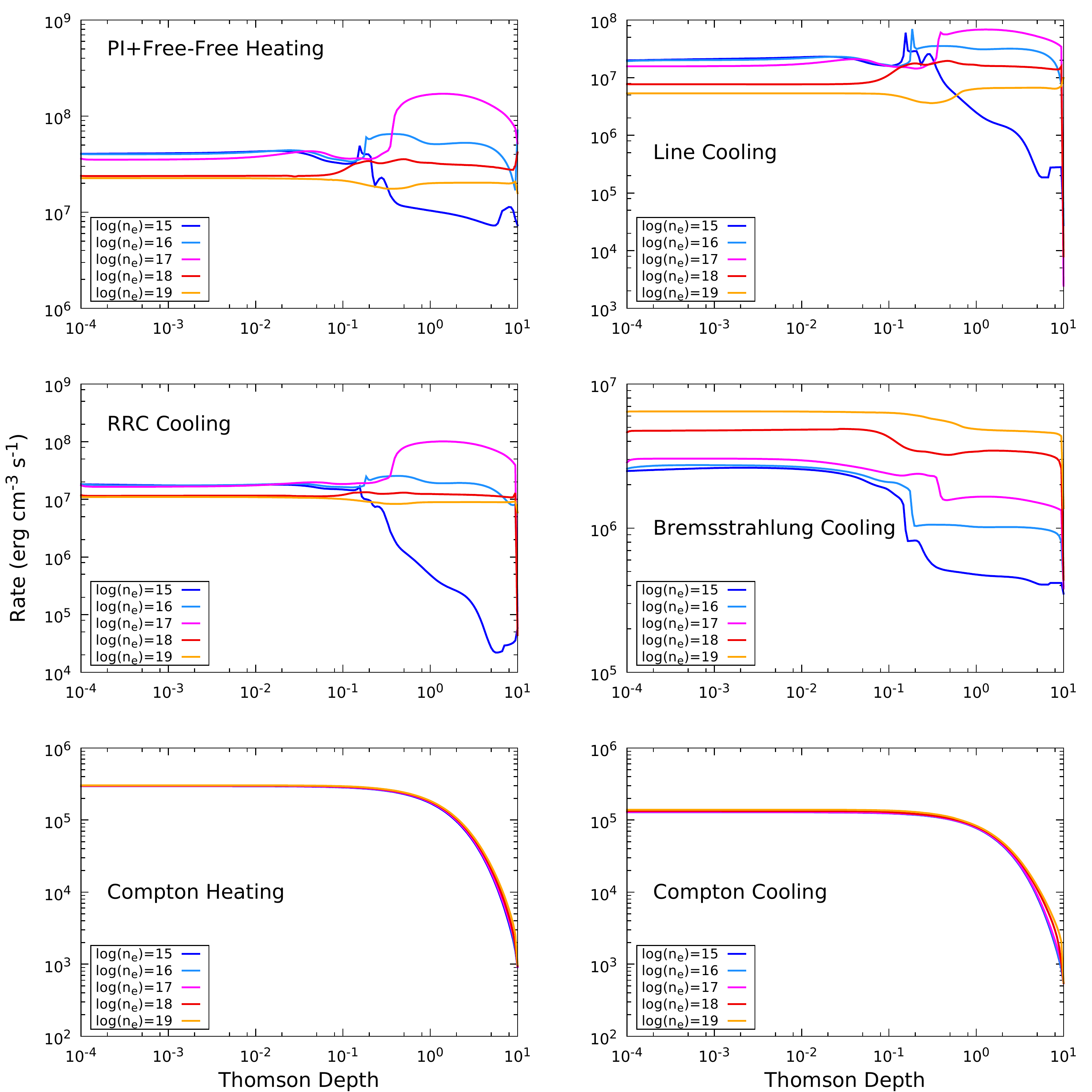}
\caption{
Various heating and cooling rates included in the reflection calculations shown
in Figure~\ref{fig:HDtemper}. Each curve is normalized in amplitude to match 
the extra factor in flux and density with respect to the $\log(n_e)=15$ case.
Thus, the curves for $\log(n_e)=16,17,18$ and $19$ are divided by $10^2,10^4,10^6$,
and $10^8$. 
}
\label{fig:rates}
\end{figure*}

All the relevant heating and cooling rates for the models mentioned above are
plotted in Figure~\ref{fig:rates}. Since we are comparing models with the same
$\xi$, a larger density implies a larger flux. A larger flux increases the
photoionization as well as the Compton heating. However, these two rates
depend linearly on density, while the free-free heating has a $n_e^2$
dependence. Compton heating and cooling scale with $n_eF_x$ independently of
the ionization or temperature, but their contribution is small compared to the
others. The Bremsstrahlung cooling has an explicit dependence with the
temperature, thus its behavior is quite similar to that seen in
Figure~\ref{fig:HDtemper}. For $\log(n_e)=15-17$, the free-free heating becomes
dominant only in the inner part of the atmosphere, at large $\tau_\mathrm{T}$.
At the surface, photoionization heating is the leading process.

%
%

%
\subsection{Reflected Spectra}\label{sec:spec}

\begin{figure*}
\centering
\includegraphics[scale=1.1,angle=0]{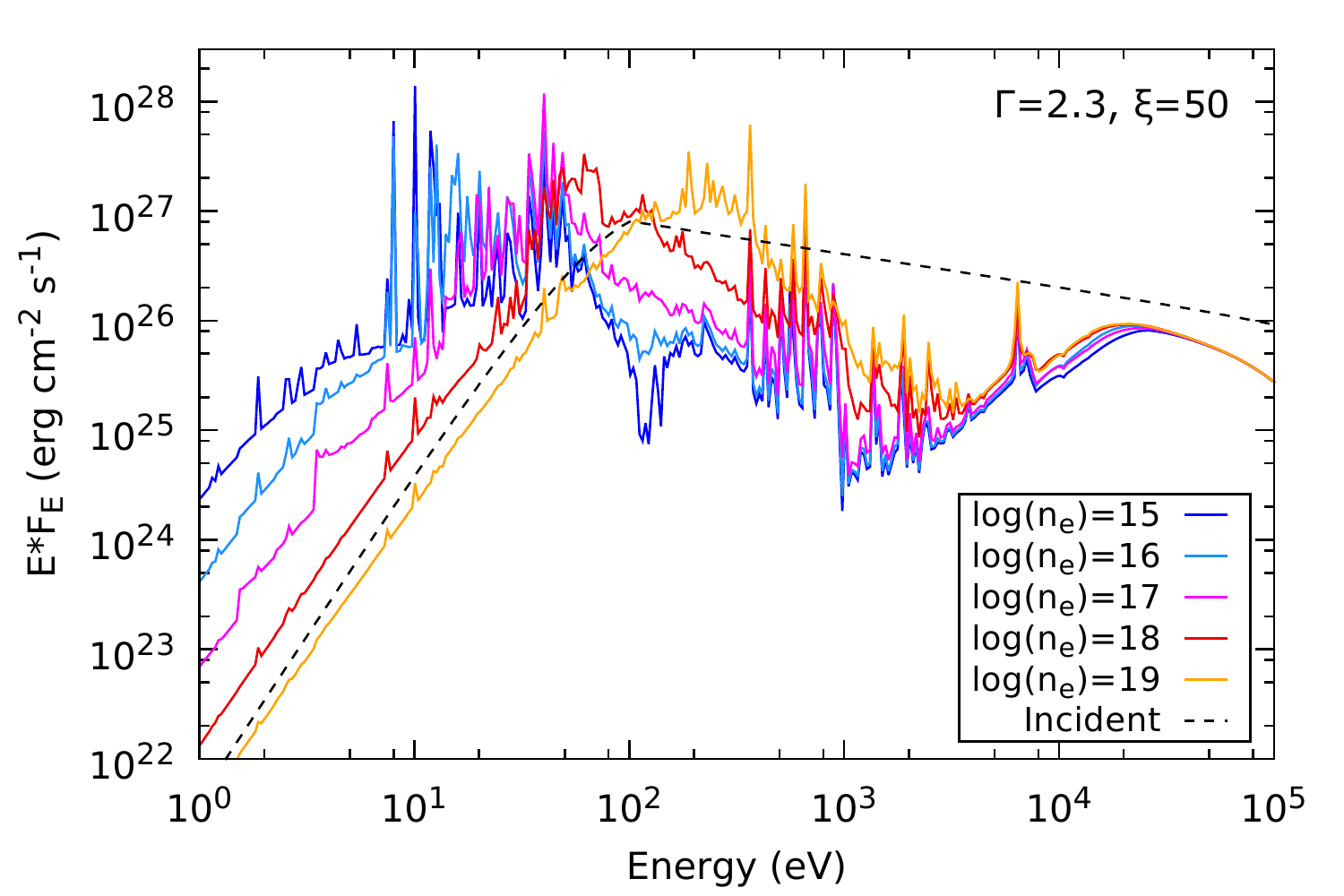}
\caption{
Angle-averaged reflected spectra for the same models shown in Figure~\ref{fig:HDtemper}.
The solid curves correspond to the density values indicated in the legend.
The dashed curve is the incident power-law spectrum that illuminates the
disk, and it is the same for all models. Each curve is normalized by an
extra factor in flux with respect to the model for $\log(n_e)=15$. This
factor is $10,10^2,10^3$ and $10^4$ for models with $\log(n_e)=16,17,18$
and $19$, respectively.
}
\label{fig:HDspectra}
\end{figure*}

The reflected spectra corresponding to the temperature solutions discussed
in Section~\ref{sec:temp} are shown in Figure~\ref{fig:HDspectra}. For simplicity
we show the angle-averaged emitted spectrum at the surface of the illuminated
disk. Each curve is normalized by an extra factor in flux with respect to the
model for $\log(n_e)=15$ such that all the continua are placed at the same level
(see Figure caption for details). The incident power-law spectrum is also shown. While
all the spectra are similar at high energies ($\ga 5$~keV), they show increasing
divergence at low energies.

The thermal part of the spectra at low energies follows the evolution
of the temperature at the illuminated zone of the atmosphere. For the three
lowest densities, there are not significant changes in the observable X-ray band
($\sim0.1-200$~keV). However, for $\log(n_e)=18$ and above the changes are
dramatic. At 1~keV there is almost 1 order of magnitude of difference in the
reflected flux. This increased emission at soft energies
agrees with the results first reported by \cite{bal04} in the context of neutron stars.
 The changes in the Fe K region are minor, although we note
the suppression of the K$\beta$ emission for the highest densities, which
indicates a higher ionization state. 

Notably, the higher density models give a soft excess which is much larger than
for low density. The main effect is that as the density rises, the
bremsstrahlung absorption in the outer layers becomes increasingly important at
lower energies. In the regions where free-free absorption is the dominant
source of opacity, the gas radiates like a blackbody at the local temperature
$T$. Thus, at low energies ($E \ll kT$) the emitted spectrum follows the
Rayleigh-Jeans law $I\propto E^2$ \citep{fel72}, causing the surface
temperature to increase.

%
%

Consider the free-free absorption coefficient for $E \ll kT$, given by
\begin{equation}\label{eq:xff}
\chi_\mathrm{ff} = 7.34\times10^{-20} n_e^2 g T^{-3/2} E^{-2},
\end{equation}
where $g$ is the temperature-averaged Gaunt factor.  The energy $E^*$ at which
the free-free equals the Thomson opacity $\chi_\mathrm{T} = n_e\sigma_\mathrm{T}$ 
(where $\sigma_\mathrm{T}=6.65\times10^{-25}$~cm$^{-2}$ is the Thomson cross section) can be
written as
\begin{equation}\label{eq:es}
E^* = \sqrt{7.34\times10^{-20} n_e g  T^{-3/2} \sigma_\mathrm{T}^{-1}},
\end{equation}
In the non-relativistic limit we may approximate $g\approx1$. Because the
temperature in the illuminated region of the slab is fairly constant, we can
evaluate the expression above for single values of density and temperature.
Looking at the region for $\tau_\mathrm{T}\la 0.01$ in the profiles of
Figure~\ref{fig:HDtemper}, we find $E^*=0.05, 0.1, 0.4, 0.8$ and $2$~keV for
$\log(n_e)=15-19$, respectively. For all energies below $E^*$,
$\chi_\mathrm{ff}$ is the dominant opacity. Note that these energies are
approximate, and in fact for energies above $\sim0.1$~keV the photoelectric
opacity from the various metals becomes important. However, this simple
derivation explains the Rayleigh-Jeans shape of the reflected spectrum at
low energies.

\begin{figure*}
\centering
\includegraphics[scale=0.4,angle=0]{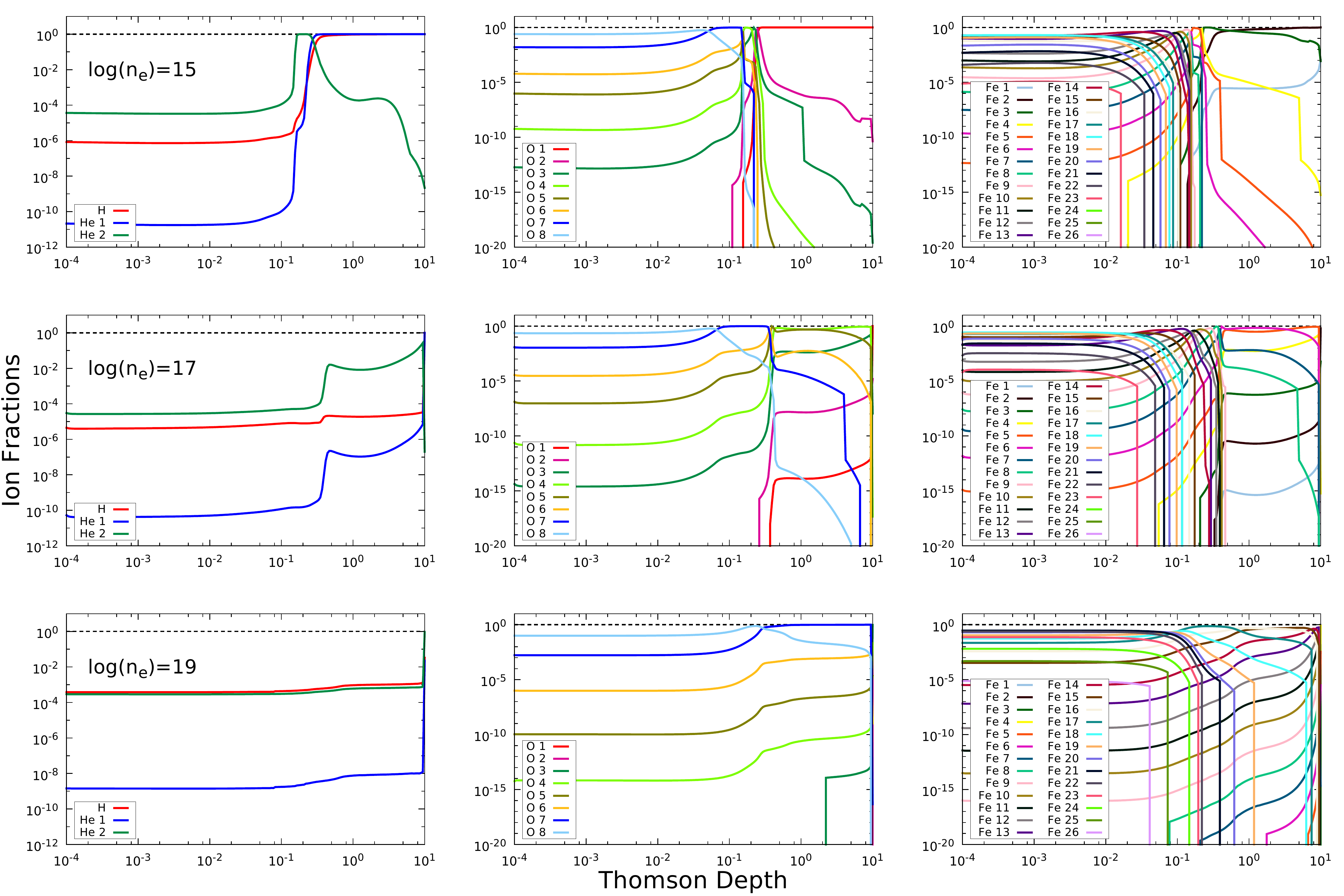}
\caption{
Fractions of the most abundant ions in the reflection calculations for $\log(n_e)=15, 17$
and $19$. Left panels show the fractions of H, He, and He~{\sc II} (labeled as He~2 in the 
figure). Middle panels show the fractions of O ions, while right panels show those for 
all Fe ions.
}
\label{fig:fractions}
\end{figure*}

The differences observed in the reflected spectra are a direct consequence of
the effects of high-density in the ionization balance of the illuminated
atmosphere. This is illustrated in Figure~\ref{fig:fractions}, where we plot
the ionic fractions for H and all ions of H, O, and Fe (the most abundant
elements), for three of the calculations presented above, namely for those with
$\log(n_e)=15, 17$, and $19$. The evolution in the ionization structure with
increasing density is quite evident.

At $\log(n_e)=15$ (upper panels of Figure~\ref{fig:fractions}), H and He are fully ionized in the illuminated
hot region near the surface, and then recombine at a small depth
($\tau_\mathrm{T}\sim 0.1$), which is the region where the temperature drops
most rapidly (Figure~\ref{fig:HDtemper}). This transition is followed by other
ions. H-like oxygen is dominant in most of the hot region, while neutral O
dominates in the colder region. Likewise, Fe~{\sc xviii} and Fe~{\sc xix} are
present near the surface, while deep inside only Fe~{\sc ii} and Fe~{\sc
iii} are observed. This structure is dramatically changed at higher densities.
For $\log(n_e)=17$, the inner regions are dominated by O~{\sc iv}, Fe~{\sc v},
and Fe~{\sc vi}. At $\log(n_e)=19$, oxygen is fully ionized in the entire slab,
while the lowest ionization stages of iron are Fe~{\sc xvii}, Fe~{\sc xvi}, and
Fe~{\sc xv}. These changes in ionic fractions affect both the photoelectric
opacity and line emission, which has direct impact on the emitted spectrum. In
general, higher density results in a hotter and more ionized atmosphere.

%
%

%
%
\section{Discussion}\label{sec:dis}

\subsection{Observational Implications for AGN}\label{sec:agn}

We have shown in Section~\ref{sec:calc} that high-density effects are very
important in the calculation of X-ray reflected spectra.
The net result is a very significant increase in the continuum flux at energies
below $\sim2$~keV. This effect could have important observational implications.
For example, the observed X-ray spectrum from many AGN often presents an emission
excess at $\sim1$~keV, referred to as the {\it soft-excess} \citep{tur88}. The
nature of this emission is still a matter of debate. Possible explanations
include Comptonized emission \citep{tit94}, reflection-reprocessed emission \citep{vas14},
or complex absorption \citep{sob07}. The calculations presented in this paper
suggest that the soft-excess could be a measure of the density in the accretion disk,
which provides a great diagnostic tool.

High-density effects are also important in extreme cases where the
illuminating source is very close to the black hole, which enhances the
reflection signal \citep{dau14}.  Recently, \cite{kar15} analyzed {\it NuSTAR}
and {\it XMM-Newton} observations of the narrow-line Seyfert~I galaxy
1H0707--495. Fits using the \relxilllp\ model (i.e., relativistic reflection
under a lampost geometry), constrained the height of the corona to be roughly
$2r_g$ (where $r_g=GM/c^2$ is the gravitational radius), and a large reflection
fraction $R_f>5.8$, which indicates a very strong illumination. However, the
ionization parameter was found to be moderate (between $\log\xi=0.3$ to
$\log\xi=2$ depending on the data set analyzed), which can only be explained by
the density being high. Using the observed luminosity and the given
ionization parameter, Kara et al. estimated a density of
$n_e\sim10^{17}$~cm$^{-3}$.

\subsection{A Simple Test on 1H0707--495}\label{sec:1h0707}

In order to test the impact of high-density reflection models in fitting
observational data, we have performed a simple analysis of one observation of
1H0707--495. We used the $\sim500$~ks exposure from the {\it XMM-Newton}
observation of the source collected between 2008 January 29 and 2008 February 6,
using only the European Photon Imaging Camera (EPIC)-pn \citep{str01}.
This is same observation analyzed by \cite{dau12}, where the details on the
data reduction are provided.

Additionally, we have produced a full grid of reflection models for
high-density atmospheres with our code \xillver. Given the very large number of
calculations required, we only produced models for a gas density of
$n_e=10^{19}$~cm$^{-3}$ (i.e., the largest value considered in the discussion of
Section~\ref{sec:calc}), with a high-energy cutoff fixed at 300~keV. The final
grid of models contains a total of 9000 spectra covering a wide range of
parameters: the slope of the illuminating power-law ($1.2 \leq \Gamma \leq
3.4$), the Fe abundance in solar units ($0.5 \leq A_\mathrm{Fe} \leq 20$), the
ionization parameter ($0 \leq \log\xi \leq 4.7$), and the inclination ($5 \leq
i \leq 89$).

The spectrum of 1H0707--495 has been previously analyzed by several authors
\citep[e.g.][]{fab09,zog10,dau12,kar15}. In general, the spectrum is found
to be dominated by reflection, with a power-law continuum much steeper than
generally found in AGN ($\Gamma \la 3$). Surprisingly, the Fe abundance is
always required to be super-solar, with values as large as 10 to 20 times 
the solar value. Most fits require either a strong thermal component (in the
form of a blackbody), or two different reflection components. Given all
these complexities, we do not seek to provide yet another detailed analysis
of this peculiar source, but rather to use it as a simple test case for 
the high-density reflection models. 

Therefore, we start by fitting the 2--10~keV energy band with a power-law
continuum plus relativistically smeared reflection (both components provided by
our model \relxill). In addition, the model includes Galactic absorption
(modeled with {\tt Tbabs}), intrinsic absorption at the redshift of the source
(modeled with {\tt zTbabs}, with $z=0.0411$). We perform two fits, one with the
standard \xillver\ grid (i.e., calculated with $n_e=10^{15}$~cm$^{-3}$), and
one with the high-density grid ($n_e=10^{19}$~cm$^{-3}$). Both models provide a
reasonably good fit for the data, with a reduced chi-square of
$\chi^2_{\nu}=1.31$ and $\chi^2_{\nu}=1.28$ for the low and high density
models, respectively, with all parameters between the two models consistent
within their uncertainties.

\begin{figure}
\centering
\includegraphics[scale=0.6,angle=0,trim={0.5cm 0 0 0}]{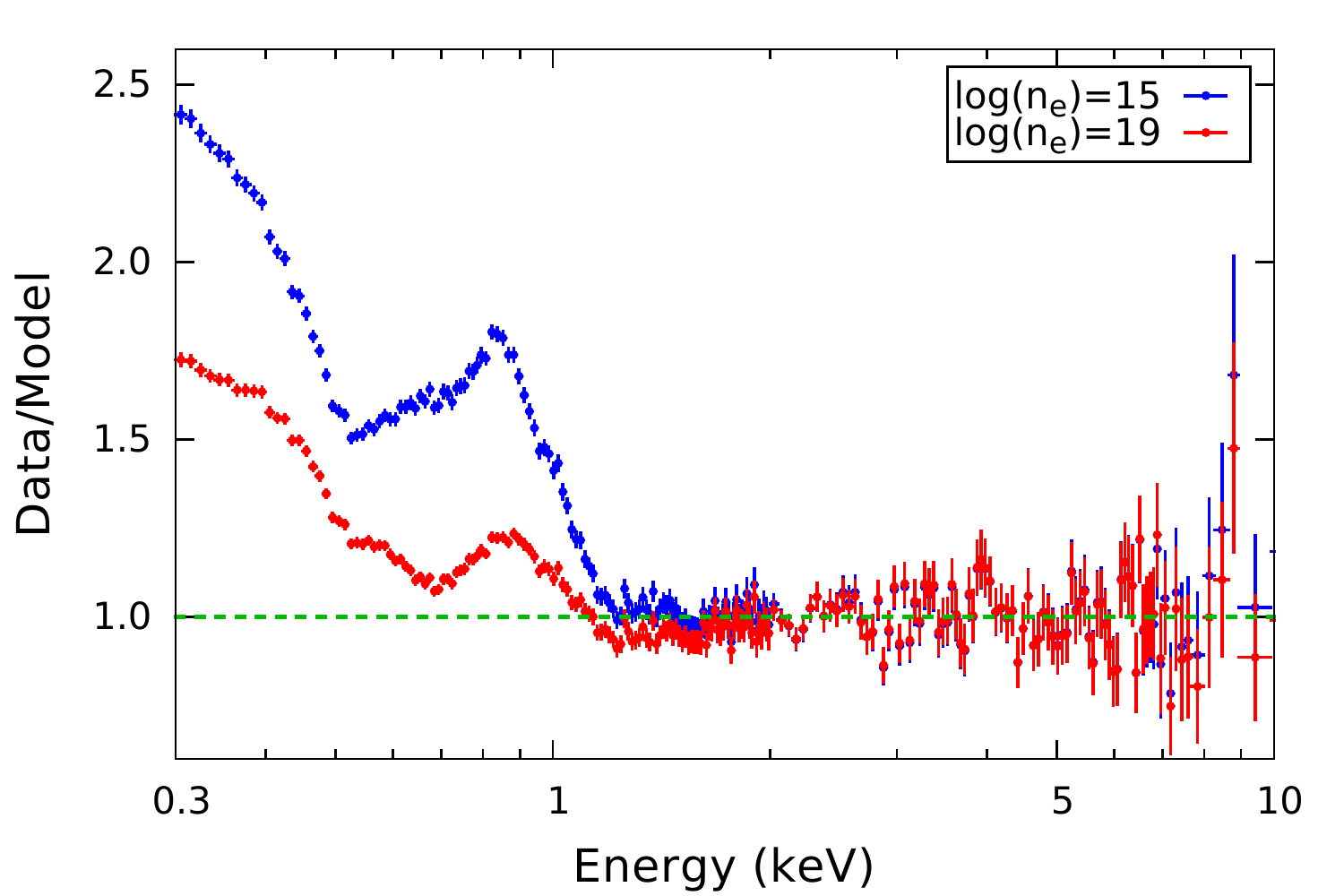}
\caption{
Ratio plots of the fits to the {\it XMM-Newton} data of 1H0707--495, using the
relativistic reflection model \relxill\ for $\log(n_e)=15$ (blue points), and
for $\log(n_e)=19$ (red points). Only the 2--10~keV band is actually fitted, while the soft
energies are included after the fit to illustrate the impact of the high-density
model in that spectral band.
}
\label{fig:HDratios}
\end{figure}

We then include the soft-energy data (i.e., the 0.3--2~keV region), {\it
without} re-fitting. Figure~\ref{fig:HDratios} shows the data-to-model ratio
plot in these two cases. Interestingly, the high-density model makes a huge
improvement in the soft band, decreasing the statistics from
$\chi^2_{\nu}=629.3$ for the low-density model, to $\chi^2_{\nu}=215.4$ for the
high-density model. Meanwhile, there is virtually no effect in the 2--10 keV
region.

An obvious question arises from this analysis: will an even higher density
improve further the fit of the soft-energies? Unfortunately, limitations in the
current atomic data prevent us from producing models at densities above
$\log(n_e)\sim18-19$. This issue is further discussed in
Section~\ref{sec:atomic}.

%
%

\subsection{Observational Implications for BHB}\label{sec:bhb}

We have discussed so far calculations appropriate for the modeling of the X-ray
reflection spectrum of AGN, i.e., steeper continua ($\Gamma\sim2$) and
relatively low ionization ($\log\xi\sim0-2$). Nevertheless, relativistic
reflection is also commonly observed in the spectra of many BHB
\citep[e.g.][]{par15,gar15,mil15}, which in general show a harder continuum 
and a higher ionization parameter ($\log\xi\ga3$). 

\begin{figure*}
\centering
\includegraphics[scale=0.6,angle=0,trim={0.5cm 0 0 0}]{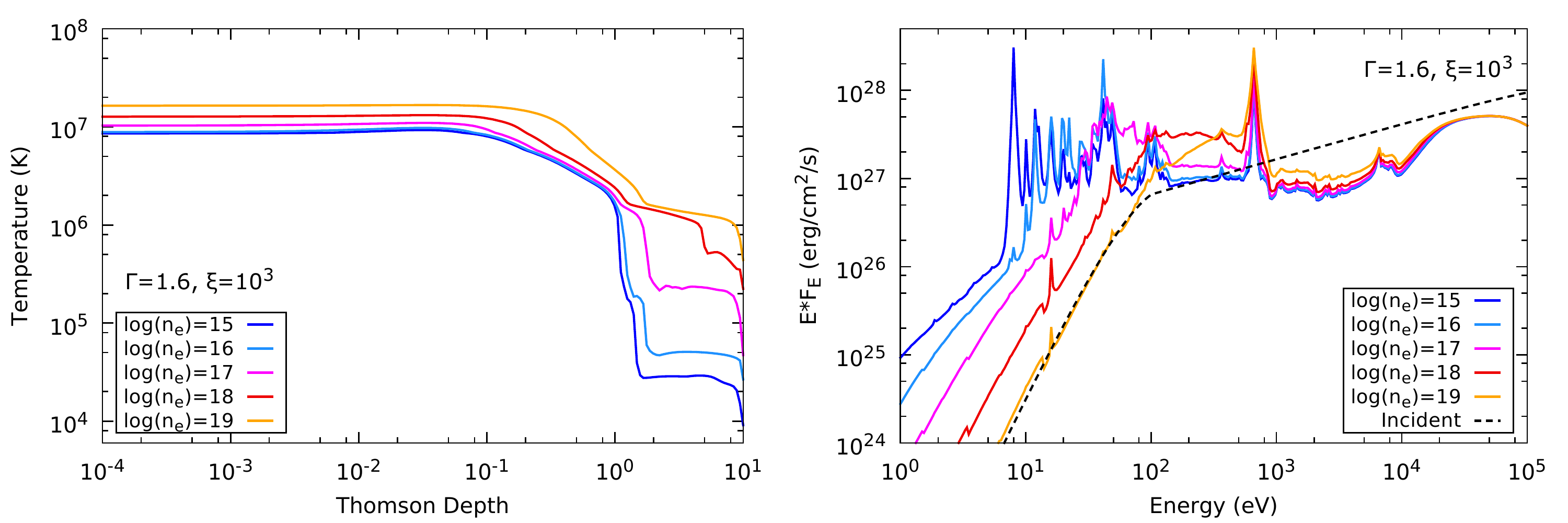}
\caption{
Reflection calculations for various densities (as indicated) using parameters 
appropriate for BHB, i.e., $\Gamma=1.6$, $\log\xi=3$, $A_\mathrm{Fe}$, and 
$E_\mathrm{cut}=300$~keV. ({\it Left}) Temperature profiles in the vertical
direction of the disk. ({\it Right}) Angle-averaged reflected spectra. The
curves are rescaled for clarity as in Figure~\ref{fig:HDspectra}.
}
\label{fig:highXi}
\end{figure*}

Figure~\ref{fig:highXi} shows the resulting temperature profiles and reflected
spectra for models with $\Gamma=1.6$ and $\xi=10^3$~erg~cm~s$^{-1}$. All other
parameters are the same as those for the models presented in
Section~\ref{sec:calc} (Figures~\ref{fig:HDtemper} and \ref{fig:HDspectra}).
The effects of high-density are less prominent at higher ionization. The
temperature near the surface remains fairly constant regardless of the density,
most likely because all elements are fully stripped and line cooling is already
suppressed (unlike the case for low $\xi$). The temperature in the inner regions
rises with density, due to the increase of the free-free heating, just as in
the low-ionization models. Once again, the most significant effect in the
reflected spectra (right panel) is the modification of the Rayleigh-Jeans
domain of the emission at soft-energies. The most prominent features are the O
Ly~$\alpha$ line at $\sim$0.8~keV, and the Fe K-shell emission at
$\sim$6.7~keV. The continuum between these two features grow with increasing
density due to the reduction of the photoelectric opacity. However, the overall
shape of the spectrum above $\sim$1~keV is very similar among all models.

%
%

In the standard $\alpha$-disk model \citep{sha73}, the midplane density of the
accretion disk around a stellar-mass black hole is a few orders of magnitude
larger than in a disk around a supermassive black hole, simply because
the density scales proportional to $m^{-1}r^{3/2}$, where $m$ is the black hole
mass and $r$ the radius. Moreover, estimates
of the gas density using multi-dimensional magneto-hydrodynamic (MHD)
simulations can produce midplane densities $\ga10^{20}$~cm$^{-3}$ for a
$10$~\msun black hole accreting at 10\% the Eddington luminosity
\citep[e.g.][]{nob10,sch13}.

In the soft and intermediate states, the power-law continuum is steep and a
strong thermal disk component dominates X-ray spectrum, which in some cases
makes the detection of the reflection signatures challenging. However, in the
hard state the thermal emission becomes faint and the spectrum is characterized
by a harder ($\Gamma<2$) power-law continuum with a high-energy cutoff. This
continuum can typically be well described by Comptonization of soft thermal
disk photons in a hot, optically-thin corona. Many sources show clear
reflection features in the hard state. 

In some cases, a weak and relatively cold disk ($T\sim 0.2$~keV) component has
been found in the hard-state spectra of some sources \citep[e.g.,
GX~339--4][]{rei08,pla15}. While the high-density effects appear to be milder
in BHB models than in those for AGN, we do not discount the possibility that
the faint soft emission observed in some BHB is due to a high-density
reflector. A detailed analysis of these new models confronted with real
observational data is required to answer this question, but outside the scope
of this paper. Moreover, as discussed above, densities even larger than ones
covered in the present calculations might be more appropriate to correctly
represent the physics near stellar-mass black holes. Such calculations are
currently infeasible due to the lack of proper atomic data (see
Section~\ref{sec:atomic}).

\subsection{Limitations of the Atomic Data}\label{sec:atomic}

The \xillver\ reflection calculations presented here make use of the
subroutines of the photoionization code \xstar\ \citep{kal01}, as well as its
atomic database \citep{bau01}.  These are used for the calculation of the
ionization and thermal balance, plus line and continuum emissivities and
opacities.  While \xstar\ incorporates the most complete atomic database
currently available for the modeling of X-ray spectra, it is also limited to
densities below $10^{18}$~cm$^{-3}$.

At high densities, electrons in bound states require less energy to liberate
them into the continuum than equivalent states in isolated atoms, because ions
see the charge from neighboring ions and electrons, which effectively depresses
the ionization potential. This effect is called {\it continuum lowering}.
Current calculations of ionization balance do not take this process into
account in a comprehensive way. In our case, \xstar\ contains a comprehensive
and accurate treatment of continuum lowering for the H- and He-like
isoelectronic sequences, and they are only applicable up to densities of
$10^{18}$~cm$^{−3}$. Plus, \xstar\ utilizes low density recombination rates
for ions from all other isoelectronic sequences, i.e. any ion with three or more
electrons.

Three-body recombination is the inverse process of collisional ionization,
and can affect all atomic levels given sufficiently high densities. It occurs when an electron
approaches an ion with kinetic energy less than the binding energy of the
recombined level, and that when second electron is within the same volume in order
to carry away the liberated energy. This process is currently not included for
all levels, but only to those excited levels most easily excited from the
ground level by electron collisions. This is supplemented by those levels which can be
populated by radiative recombination. As three-body recombination occurs
preferentially to highly excited levels, the line emission from these levels
can be strongly enhanced at high-densities.

The emission of X-ray lines depends on the rates from many processes,
including photoionization, radiative and dielectronic recombination, electron
impact collisional excitation and ionization, Auger decay and autoionization,
and bound-bound radiative decay. Plasma density can affect all of these, in
principle, via effects of screening the nuclear charge on atomic structure.
This can lead to rate coefficients which can be either larger or smaller than
the low density rates.

Stimulated processes will enhance the rates for recombination and radiative
decay at high-densities, and it is straightforward to include in the
calculation of level populations. However, so far \xstar\ only includes this
for recombination, given the current limitation to densities below
$10^{18}$~cm$^{−3}$ (i.e., stimulated processes become important at larger
densities).

%
%
\section{Conclusions}\label{sec:conc}

We have explored the effects of high densities on models of photoionized
plasmas and reflection spectra. The primary effect of increasing the
density is the suppression of line cooling near the disk's surface, and
the increase of free-free heating in the deeper regions, which results
in an increase of the gas temperature. These effects in turn affect the
ionization balance, typically increasing the ionization state of the gas
as the density increases. The most obvious effect on the reflection
spectrum is the increase of the thermal emission at
low energies, which follows a Rayleigh-Jeans law. Additionally, the
higher temperature also results in a higher ionization state, thereby
affecting the emission line spectrum and continuum photoelectric
absorption.

In the context of AGN, the enhanced emission at low energies could
explain to some degree the soft excess observed in many sources. Higher
densities might also explain the blackbody component that is required in
fits to the spectra of extreme narrow-line Seyfert I AGN such as
1H0707--495 \citep{fab09} and IRAS13224--3809 \citep{pon10}, and
possibly also fits to data indicating the presence of a second reflector
with higher ionization identified by \cite{dau12}. Generally, the
effects of high density are more dramatic for low ionization-state
models.

High density models are also important when the illumination of the
inner disk is very strong. Specifically, for bright AGN, with the
illuminating source near the black hole and moderate ionization, the
density of the reflector is expected to exceed the typical densities
employed in standard reflection models \citep[e.g.,][]{par14,kar15}.

As a first step, using our code \xillver\ we have produced a new library
of reflection spectra that are publicly
available\footnote{\url{http://hea-www.cfa.harvard.edu/~javier/xillver/}}. The
grid of models is for a single high value of the density,
$n_e=10^{19}$~cm$^{-3}$, while covering wide ranges of all the other
relevant parameters (except for $E_\mathrm{cut}$, which is fixed at
300~keV). This limited library of models is provided to enable
individuals to test whether a high-density model of a peculiar source
provides a better fit to the data than conventional reflection
models. Ultimately, a large effort will be required to produce a
comprehensive set of models that accurately describe the microphysics
associated with the effects of high density. To a degree, we expect the
density to be degenerate with the ionization parameter at some energies,
though it may be possible to use the soft excess as a discriminator.

The atomic data currently available is somewhat incomplete and
applicable only to plasma densities less than
$n_e=10^{19}$~cm$^{-3}$. This density limit imposes a serious limitation
on our reflection models since the densities in BHB accretion disks are
estimated to be significantly higher. Similarly, in other important
environments the densities are thought to be $>n_e=10^{19}$~cm$^{-3}$,
such as the accretion disks of neutron stars, the atmospheres of white
dwarfs, and partially ionized outflows from galactic sources
\citep{mil15b}. At such densities, much of the rate data in current
atomic databases is inapplicable and new atomic calculations are sorely
needed. 

A principal use of reflection models is in estimating the spins of black
holes where the fits to data are driven by the profile of the Fe K
emission line and edge. Fits to the spectra of many sources have shown a
clear correlation between the spin of the black hole and the abundance
of Fe \citep{rey12,ste12}. Moreover, the abundances are typically
significantly greater than the solar value. The AGN 1H0707--495 is the
most dramatic case with $A_\mathrm{Fe}\sim10-20$ \citep{fab09,dau12}.
Meanwhile, super-solar abundances are also inferred in fitting
high-quality spectra of BHBs such as Cyg~X-1 \citep{par15} and
GX~339--4 \citep{fue15,gar15}. 

Because no concrete physical mechanism for extreme Fe enrichment has
been identified, it appears that there is a deficiency in the models. In
current reflection models, it is the Fe-abundance parameter alone that
controls the strength of the line emission. A possible explanation for
the super-solar abundances is that current moderate-density models fail
to include a significant line-production mechanism. It is imperative to
explore this possibility by identifying and accurately including
high-density plasma effects in computing atomic data, and then
implementing these data in a new generation of reflection models.

The development of accurate high-density reflection models is now a
priority given the broadband, high-resolution, pile-up-free data that
{\it NuSTAR} and the newly-launched mission {\it Hitomi} \citep[{\it
  Astro-H};][]{tak10} are providing, and that {\it NICER} \citep{gen12}
will soon provide. The latter two missions furthermore are sensitive down
to $\sim0.2$~keV, making their data ideal in searching for the predicted
signatures of high-density plasma effects.

%
%
%
\vspace{0.25in}

JG and JEM acknowledge the support of a CGPS grant from the Smithsonian Institution.
ACF acknowledges ERC Advanced Grant 340442 Feedback.
JFS has been supported by the Einstein Fellowship grant PF5-160144.
%
%
%
\bibliographystyle{mnras}
\bibliography{my-references}

\begin{thebibliography}{}
\makeatletter
\relax
\def\mn@urlcharsother{\let\do\@makeother \do\$\do\&\do\#\do\^\do\_\do\%\do\~}
\def\mn@doi{\begingroup\mn@urlcharsother \@ifnextchar [ {\mn@doi@}
  {\mn@doi@[]}}
\def\mn@doi@[#1]#2{\def\@tempa{#1}\ifx\@tempa\@empty \href
  {http://dx.doi.org/#2} {doi:#2}\else \href {http://dx.doi.org/#2} {#1}\fi
  \endgroup}
\def\mn@eprint#1#2{\mn@eprint@#1:#2::\@nil}
\def\mn@eprint@arXiv#1{\href {http://arxiv.org/abs/#1} {{\tt arXiv:#1}}}
\def\mn@eprint@dblp#1{\href {http://dblp.uni-trier.de/rec/bibtex/#1.xml}
  {dblp:#1}}
\def\mn@eprint@#1:#2:#3:#4\@nil{\def\@tempa {#1}\def\@tempb {#2}\def\@tempc
  {#3}\ifx \@tempc \@empty \let \@tempc \@tempb \let \@tempb \@tempa \fi \ifx
  \@tempb \@empty \def\@tempb {arXiv}\fi \@ifundefined
  {mn@eprint@\@tempb}{\@tempb:\@tempc}{\expandafter \expandafter \csname
  mn@eprint@\@tempb\endcsname \expandafter{\@tempc}}}

\bibitem[\protect\citeauthoryear{{Ballantyne}}{{Ballantyne}}{2004}]{bal04}
{Ballantyne} D.~R.,  2004, \mn@doi [\mnras] {10.1111/j.1365-2966.2004.07767.x},
  \href {http://adsabs.harvard.edu/abs/2004MNRAS.351...57B} {351, 57}

\bibitem[\protect\citeauthoryear{{Ballantyne}, {Ross}  \&
  {Fabian}}{{Ballantyne} et~al.}{2001}]{bal01}
{Ballantyne} D.~R.,  {Ross} R.~R.,   {Fabian} A.~C.,  2001, \mn@doi [\mnras]
  {10.1046/j.1365-8711.2001.04432.x}, \href
  {http://adsabs.harvard.edu/abs/2001MNRAS.327...10B} {327, 10}

\bibitem[\protect\citeauthoryear{{Bautista} \& {Kallman}}{{Bautista} \&
  {Kallman}}{2001}]{bau01}
{Bautista} M.~A.,  {Kallman} T.~R.,  2001, \mn@doi [\apjs] {10.1086/320363},
  \href {http://adsabs.harvard.edu/abs/2001ApJS..134..139B} {134, 139}

\bibitem[\protect\citeauthoryear{{Cackett} et~al.,}{{Cackett}
  et~al.}{2010}]{cak10}
{Cackett} E.~M.,  et~al., 2010, \mn@doi [\apj] {10.1088/0004-637X/720/1/205},
  \href {http://adsabs.harvard.edu/abs/2010ApJ...720..205C} {720, 205}

\bibitem[\protect\citeauthoryear{{Chen}, {Gou}, {McClintock}, {Steiner}, {Wu},
  {Xu}, {Orosz}  \& {Xiang}}{{Chen} et~al.}{2015}]{che15}
{Chen} Z.,  {Gou} L.,  {McClintock} J.~E.,  {Steiner} J.~F.,  {Wu} J.,  {Xu}
  W.,  {Orosz} J.,   {Xiang} Y.,  2015, preprint, \href
  {http://adsabs.harvard.edu/abs/2016arXiv160100615C} {} (\mn@eprint {arXiv}
  {1601.00615})

\bibitem[\protect\citeauthoryear{{Dauser}, {Wilms}, {Reynolds}  \&
  {Brenneman}}{{Dauser} et~al.}{2010}]{dau10}
{Dauser} T.,  {Wilms} J.,  {Reynolds} C.~S.,   {Brenneman} L.~W.,  2010,
  \mn@doi [\mnras] {10.1111/j.1365-2966.2010.17393.x}, \href
  {http://adsabs.harvard.edu/abs/2010MNRAS.409.1534D} {409, 1534}

\bibitem[\protect\citeauthoryear{{Dauser} et~al.,}{{Dauser}
  et~al.}{2012}]{dau12}
{Dauser} T.,  et~al., 2012, \mnras, 422, 1914

\bibitem[\protect\citeauthoryear{{Dauser}, {Garcia}, {Wilms}, {B{\"o}ck},
  {Brenneman}, {Falanga}, {Fukumura}  \& {Reynolds}}{{Dauser}
  et~al.}{2013}]{dau13}
{Dauser} T.,  {Garcia} J.,  {Wilms} J.,  {B{\"o}ck} M.,  {Brenneman} L.~W.,
  {Falanga} M.,  {Fukumura} K.,   {Reynolds} C.~S.,  2013, \mn@doi [\mnras]
  {10.1093/mnras/sts710}, \href
  {http://adsabs.harvard.edu/abs/2013MNRAS.430.1694D} {430, 1694}

\bibitem[\protect\citeauthoryear{{Dauser}, {Garc{\'{\i}}a}, {Parker}, {Fabian}
  \& {Wilms}}{{Dauser} et~al.}{2014}]{dau14}
{Dauser} T.,  {Garc{\'{\i}}a} J.,  {Parker} M.~L.,  {Fabian} A.~C.,   {Wilms}
  J.,  2014, \mn@doi [\mnras] {10.1093/mnrasl/slu125}, \href
  {http://adsabs.harvard.edu/abs/2014MNRAS.444L.100D} {444, L100}

\bibitem[\protect\citeauthoryear{{Dov{\v c}iak}, {Karas}  \& {Yaqoob}}{{Dov{\v
  c}iak} et~al.}{2004}]{dov04}
{Dov{\v c}iak} M.,  {Karas} V.,   {Yaqoob} T.,  2004, \mn@doi [\apjs]
  {10.1086/421115}, \href {http://adsabs.harvard.edu/abs/2004ApJS..153..205D}
  {153, 205}

\bibitem[\protect\citeauthoryear{{Fabian} \& {Ross}}{{Fabian} \&
  {Ross}}{2010}]{fab10}
{Fabian} A.~C.,  {Ross} R.~R.,  2010, \mn@doi [\ssr]
  {10.1007/s11214-010-9699-y}, \href
  {http://adsabs.harvard.edu/abs/2010SSRv..157..167F} {157, 167}

\bibitem[\protect\citeauthoryear{{Fabian}, {Iwasawa}, {Reynolds}  \&
  {Young}}{{Fabian} et~al.}{2000}]{fab00}
{Fabian} A.~C.,  {Iwasawa} K.,  {Reynolds} C.~S.,   {Young} A.~J.,  2000,
  \mn@doi [\pasp] {10.1086/316610}, \href
  {http://adsabs.harvard.edu/abs/2000PASP..112.1145F} {112, 1145}

\bibitem[\protect\citeauthoryear{{Fabian} et~al.,}{{Fabian}
  et~al.}{2009}]{fab09}
{Fabian} A.~C.,  et~al., 2009, \mn@doi [\nat] {10.1038/nature08007}, \href
  {http://adsabs.harvard.edu/abs/2009Natur.459..540F} {459, 540}

\bibitem[\protect\citeauthoryear{{Felten} \& {Rees}}{{Felten} \&
  {Rees}}{1972}]{fel72}
{Felten} J.~E.,  {Rees} M.~J.,  1972, \aap, \href
  {http://adsabs.harvard.edu/abs/1972A%26A....17..226F} {17, 226}

\bibitem[\protect\citeauthoryear{{Fuerst} et~al.,}{{Fuerst}
  et~al.}{2015}]{fue15}
{Fuerst} F.,  et~al., 2015, preprint, \href
  {http://adsabs.harvard.edu/abs/2015arXiv150601381F} {} (\mn@eprint {arXiv}
  {1506.01381})

\bibitem[\protect\citeauthoryear{{Garc{\'{\i}}a} \& {Kallman}}{{Garc{\'{\i}}a}
  \& {Kallman}}{2010}]{gar10}
{Garc{\'{\i}}a} J.,  {Kallman} T.~R.,  2010, \mn@doi [\apj]
  {10.1088/0004-637X/718/2/695}, \href
  {http://adsabs.harvard.edu/abs/2010ApJ...718..695G} {718, 695}

\bibitem[\protect\citeauthoryear{{Garc{\'{\i}}a}, {Mendoza}, {Bautista},
  {Gorczyca}, {Kallman}  \& {Palmeri}}{{Garc{\'{\i}}a} et~al.}{2005}]{gar05}
{Garc{\'{\i}}a} J.,  {Mendoza} C.,  {Bautista} M.~A.,  {Gorczyca} T.~W.,
  {Kallman} T.~R.,   {Palmeri} P.,  2005, \mn@doi [\apjs] {10.1086/428712},
  \href {http://adsabs.harvard.edu/abs/2005ApJS..158...68G} {158, 68}

\bibitem[\protect\citeauthoryear{{Garc{\'{\i}}a} et~al.,}{{Garc{\'{\i}}a}
  et~al.}{2009}]{gar09}
{Garc{\'{\i}}a} J.,  et~al., 2009, \mn@doi [\apjs]
  {10.1088/0067-0049/185/2/477}, \href
  {http://adsabs.harvard.edu/abs/2009ApJS..185..477G} {185, 477}

\bibitem[\protect\citeauthoryear{{Garc{\'{\i}}a}, {Dauser}, {Reynolds},
  {Kallman}, {McClintock}, {Wilms}  \& {Eikmann}}{{Garc{\'{\i}}a}
  et~al.}{2013}]{gar13a}
{Garc{\'{\i}}a} J.,  {Dauser} T.,  {Reynolds} C.~S.,  {Kallman} T.~R.,
  {McClintock} J.~E.,  {Wilms} J.,   {Eikmann} W.,  2013, \mn@doi [\apj]
  {10.1088/0004-637X/768/2/146}, \href
  {http://adsabs.harvard.edu/abs/2013ApJ...768..146G} {768, 146}

\bibitem[\protect\citeauthoryear{{Garc{\'{\i}}a} et~al.,}{{Garc{\'{\i}}a}
  et~al.}{2014}]{gar14a}
{Garc{\'{\i}}a} J.,  et~al., 2014, \mn@doi [\apj] {10.1088/0004-637X/782/2/76},
  \href {http://cdsads.u-strasbg.fr/abs/2014ApJ...782...76G} {782, 76}

\bibitem[\protect\citeauthoryear{{Garc\'ia}, {Steiner}, {McClintock},
  {Remillard}, {Grinberg}  \& {Dauser}}{{Garc\'ia} et~al.}{2015}]{gar15}
{Garc\'ia} J.~A.,  {Steiner} J.~F.,  {McClintock} J.~E.,  {Remillard} R.~A.,
  {Grinberg} V.,   {Dauser} T.,  2015, \mn@doi [\apj]
  {10.1088/0004-637X/813/2/84}, \href
  {http://adsabs.harvard.edu/abs/2015ApJ...813...84G} {813, 84}

\bibitem[\protect\citeauthoryear{{Gendreau}, {Arzoumanian}  \&
  {Okajima}}{{Gendreau} et~al.}{2012}]{gen12}
{Gendreau} K.~C.,  {Arzoumanian} Z.,   {Okajima} T.,  2012, in Space Telescopes
  and Instrumentation 2012: Ultraviolet to Gamma Ray. p. 844313,
  \mn@doi{10.1117/12.926396}

\bibitem[\protect\citeauthoryear{{Gou} et~al.,}{{Gou} et~al.}{2014}]{gou14}
{Gou} L.,  et~al., 2014, \mn@doi [\apj] {10.1088/0004-637X/790/1/29}, \href
  {http://adsabs.harvard.edu/abs/2014ApJ...790...29G} {790, 29}

\bibitem[\protect\citeauthoryear{{Kallman} \& {Bautista}}{{Kallman} \&
  {Bautista}}{2001}]{kal01}
{Kallman} T.,  {Bautista} M.,  2001, \mn@doi [\apjs] {10.1086/319184}, \href
  {http://adsabs.harvard.edu/abs/2001ApJS..133..221K} {133, 221}

\bibitem[\protect\citeauthoryear{{Kallman}, {Palmeri}, {Bautista}, {Mendoza}
  \& {Krolik}}{{Kallman} et~al.}{2004}]{kal04}
{Kallman} T.~R.,  {Palmeri} P.,  {Bautista} M.~A.,  {Mendoza} C.,   {Krolik}
  J.~H.,  2004, \mn@doi [\apjs] {10.1086/424039}, \href
  {http://adsabs.harvard.edu/cgi-bin/nph-bib_query?bibcode=2004ApJS..155..675K&db_key=AST}
  {155, 675}

\bibitem[\protect\citeauthoryear{{Kara} et~al.,}{{Kara} et~al.}{2015}]{kar15}
{Kara} E.,  et~al., 2015, \mn@doi [\mnras] {10.1093/mnras/stu2136}, \href
  {http://adsabs.harvard.edu/abs/2015MNRAS.446..737K} {446, 737}

\bibitem[\protect\citeauthoryear{{Madej}, {Garc{\'{\i}}a}, {Jonker}, {Parker},
  {Ross}, {Fabian}  \& {Chenevez}}{{Madej} et~al.}{2014}]{mad14}
{Madej} O.~K.,  {Garc{\'{\i}}a} J.,  {Jonker} P.~G.,  {Parker} M.~L.,  {Ross}
  R.,  {Fabian} A.~C.,   {Chenevez} J.,  2014, \mn@doi [\mnras]
  {10.1093/mnras/stu884}, \href
  {http://adsabs.harvard.edu/abs/2014MNRAS.442.1157M} {442, 1157}

\bibitem[\protect\citeauthoryear{{McClintock}, {Narayan}  \&
  {Steiner}}{{McClintock} et~al.}{2014}]{mcc14}
{McClintock} J.~E.,  {Narayan} R.,   {Steiner} J.~F.,  2014, \mn@doi [\ssr]
  {10.1007/s11214-013-0003-9}, \href
  {http://adsabs.harvard.edu/abs/2014SSRv..183..295M} {183, 295}

\bibitem[\protect\citeauthoryear{{Mihalas}}{{Mihalas}}{1978}]{mih78}
{Mihalas} D.,  1978, {Stellar atmospheres}.
2nd ed.; San Francisco, CA: Freeman

\bibitem[\protect\citeauthoryear{{Miller} \& {Miller}}{{Miller} \&
  {Miller}}{2015}]{mil15}
{Miller} M.~C.,  {Miller} J.~M.,  2015, \mn@doi [\physrep]
  {10.1016/j.physrep.2014.09.003}, \href
  {http://adsabs.harvard.edu/abs/2015PhR...548....1M} {548, 1}

\bibitem[\protect\citeauthoryear{{Miller}, {Fabian}, {Kaastra}, {Kallman},
  {King}, {Proga}, {Raymond}  \& {Reynolds}}{{Miller} et~al.}{2015}]{mil15b}
{Miller} J.~M.,  {Fabian} A.~C.,  {Kaastra} J.,  {Kallman} T.,  {King} A.~L.,
  {Proga} D.,  {Raymond} J.,   {Reynolds} C.~S.,  2015, \mn@doi [\apj]
  {10.1088/0004-637X/814/2/87}, \href
  {http://adsabs.harvard.edu/abs/2015ApJ...814...87M} {814, 87}

\bibitem[\protect\citeauthoryear{{Mukai}, {Rana}, {Bernardini}  \& {de
  Martino}}{{Mukai} et~al.}{2015}]{muk15}
{Mukai} K.,  {Rana} V.,  {Bernardini} F.,   {de Martino} D.,  2015, \mn@doi
  [\apjl] {10.1088/2041-8205/807/2/L30}, \href
  {http://adsabs.harvard.edu/abs/2015ApJ...807L..30M} {807, L30}

\bibitem[\protect\citeauthoryear{{Nayakshin} \& {Kallman}}{{Nayakshin} \&
  {Kallman}}{2001}]{nay01}
{Nayakshin} S.,  {Kallman} T.~R.,  2001, \mn@doi [\apj] {10.1086/318250}, \href
  {http://adsabs.harvard.edu/abs/2001ApJ...546..406N} {546, 406}

\bibitem[\protect\citeauthoryear{{Noble}, {Krolik}  \& {Hawley}}{{Noble}
  et~al.}{2010}]{nob10}
{Noble} S.~C.,  {Krolik} J.~H.,   {Hawley} J.~F.,  2010, \mn@doi [\apj]
  {10.1088/0004-637X/711/2/959}, \href
  {http://adsabs.harvard.edu/abs/2010ApJ...711..959N} {711, 959}

\bibitem[\protect\citeauthoryear{{Parker} et~al.,}{{Parker}
  et~al.}{2014}]{par14}
{Parker} M.~L.,  et~al., 2014, \mn@doi [\mnras] {10.1093/mnras/stu1246}, \href
  {http://adsabs.harvard.edu/abs/2014MNRAS.443.1723P} {443, 1723}

\bibitem[\protect\citeauthoryear{{Parker} et~al.,}{{Parker}
  et~al.}{2015}]{par15}
{Parker} M.~L.,  et~al., 2015, \mn@doi [\apj] {10.1088/0004-637X/808/1/9},
  \href {http://adsabs.harvard.edu/abs/2015ApJ...808....9P} {808, 9}

\bibitem[\protect\citeauthoryear{{Patrick}, {Reeves}, {Lobban}, {Porquet}  \&
  {Markowitz}}{{Patrick} et~al.}{2011}]{pat11}
{Patrick} A.~R.,  {Reeves} J.~N.,  {Lobban} A.~P.,  {Porquet} D.,   {Markowitz}
  A.~G.,  2011, \mn@doi [\mnras] {10.1111/j.1365-2966.2011.19224.x}, \href
  {http://adsabs.harvard.edu/abs/2011MNRAS.416.2725P} {416, 2725}

\bibitem[\protect\citeauthoryear{{Plant}, {Fender}, {Ponti}, {Mu{\~n}oz-Darias}
   \& {Coriat}}{{Plant} et~al.}{2015}]{pla15}
{Plant} D.~S.,  {Fender} R.~P.,  {Ponti} G.,  {Mu{\~n}oz-Darias} T.,   {Coriat}
  M.,  2015, \mn@doi [\aap] {10.1051/0004-6361/201423925}, \href
  {http://adsabs.harvard.edu/abs/2015A%26A...573A.120P} {573, A120}

\bibitem[\protect\citeauthoryear{{Ponti} et~al.,}{{Ponti} et~al.}{2010}]{pon10}
{Ponti} G.,  et~al., 2010, \mn@doi [\mnras] {10.1111/j.1365-2966.2010.16852.x},
  \href {http://adsabs.harvard.edu/abs/2010MNRAS.406.2591P} {406, 2591}

\bibitem[\protect\citeauthoryear{{Reis}, {Fabian}, {Ross}, {Miniutti}, {Miller}
   \& {Reynolds}}{{Reis} et~al.}{2008}]{rei08}
{Reis} R.~C.,  {Fabian} A.~C.,  {Ross} R.~R.,  {Miniutti} G.,  {Miller} J.~M.,
   {Reynolds} C.,  2008, \mn@doi [\mnras] {10.1111/j.1365-2966.2008.13358.x},
  \href {http://adsabs.harvard.edu/abs/2008MNRAS.387.1489R} {387, 1489}

\bibitem[\protect\citeauthoryear{{Reynolds}}{{Reynolds}}{2014}]{rey14}
{Reynolds} C.~S.,  2014, \mn@doi [\ssr] {10.1007/s11214-013-0006-6}, \href
  {http://adsabs.harvard.edu/abs/2014SSRv..183..277R} {183, 277}

\bibitem[\protect\citeauthoryear{{Reynolds} \& {Nowak}}{{Reynolds} \&
  {Nowak}}{2003}]{rey03}
{Reynolds} C.~S.,  {Nowak} M.~A.,  2003, \mn@doi [\physrep]
  {10.1016/S0370-1573(02)00584-7}, \href
  {http://adsabs.harvard.edu/abs/2003PhR...377..389R} {377, 389}

\bibitem[\protect\citeauthoryear{{Reynolds}, {Brenneman}, {Lohfink}, {Trippe},
  {Miller}, {Fabian}  \& {Nowak}}{{Reynolds} et~al.}{2012}]{rey12}
{Reynolds} C.~S.,  {Brenneman} L.~W.,  {Lohfink} A.~M.,  {Trippe} M.~L.,
  {Miller} J.~M.,  {Fabian} A.~C.,   {Nowak} M.~A.,  2012, \mn@doi [\apj]
  {10.1088/0004-637X/755/2/88}, \href
  {http://adsabs.harvard.edu/abs/2012ApJ...755...88R} {755, 88}

\bibitem[\protect\citeauthoryear{{Ross} \& {Fabian}}{{Ross} \&
  {Fabian}}{1993}]{ros93}
{Ross} R.~R.,  {Fabian} A.~C.,  1993, \mnras, \href
  {http://adsabs.harvard.edu/cgi-bin/nph-bib_query?bibcode=1993MNRAS.261...74R&db_key=AST}
  {261, 74}

\bibitem[\protect\citeauthoryear{{Ross} \& {Fabian}}{{Ross} \&
  {Fabian}}{2005}]{ros05}
{Ross} R.~R.,  {Fabian} A.~C.,  2005, \mn@doi [\mnras]
  {10.1111/j.1365-2966.2005.08797.x}, \href
  {http://adsabs.harvard.edu/abs/2005MNRAS.358..211R} {358, 211}

\bibitem[\protect\citeauthoryear{{Ross} \& {Fabian}}{{Ross} \&
  {Fabian}}{2007}]{ros07}
{Ross} R.~R.,  {Fabian} A.~C.,  2007, \mn@doi [\mnras]
  {10.1111/j.1365-2966.2007.12339.x}, \href
  {http://adsabs.harvard.edu/abs/2007MNRAS.381.1697R} {381, 1697}

\bibitem[\protect\citeauthoryear{{R{\'o}{\.z}a{\'n}ska} \&
  {Madej}}{{R{\'o}{\.z}a{\'n}ska} \& {Madej}}{2008}]{roz08}
{R{\'o}{\.z}a{\'n}ska} A.,  {Madej} J.,  2008, \mn@doi [\mnras]
  {10.1111/j.1365-2966.2008.13173.x}, \href
  {http://adsabs.harvard.edu/abs/2008MNRAS.386.1872R} {386, 1872}

\bibitem[\protect\citeauthoryear{{R{\'o}{\.z}a{\'n}ska}, {Dumont}, {Czerny}  \&
  {Collin}}{{R{\'o}{\.z}a{\'n}ska} et~al.}{2002}]{roz02}
{R{\'o}{\.z}a{\'n}ska} A.,  {Dumont} A.-M.,  {Czerny} B.,   {Collin} S.,  2002,
  \mn@doi [\mnras] {10.1046/j.1365-8711.2002.05338.x}, \href
  {http://adsabs.harvard.edu/abs/2002MNRAS.332..799R} {332, 799}

\bibitem[\protect\citeauthoryear{{Schnittman}, {Krolik}  \&
  {Noble}}{{Schnittman} et~al.}{2013}]{sch13}
{Schnittman} J.~D.,  {Krolik} J.~H.,   {Noble} S.~C.,  2013, \mn@doi [\apj]
  {10.1088/0004-637X/769/2/156}, \href
  {http://adsabs.harvard.edu/abs/2013ApJ...769..156S} {769, 156}

\bibitem[\protect\citeauthoryear{{Shakura} \& {Sunyaev}}{{Shakura} \&
  {Sunyaev}}{1973}]{sha73}
{Shakura} N.~I.,  {Sunyaev} R.~A.,  1973, \aap, \href
  {http://adsabs.harvard.edu/cgi-bin/nph-bib_query?bibcode=1973A%26A....24..337S&db_key=AST}
  {24, 337}

\bibitem[\protect\citeauthoryear{{Sobolewska} \& {Done}}{{Sobolewska} \&
  {Done}}{2007}]{sob07}
{Sobolewska} M.~A.,  {Done} C.,  2007, \mn@doi [\mnras]
  {10.1111/j.1365-2966.2006.11117.x}, \href
  {http://adsabs.harvard.edu/abs/2007MNRAS.374..150S} {374, 150}

\bibitem[\protect\citeauthoryear{{Steiner} \& {McClintock}}{{Steiner} \&
  {McClintock}}{2012}]{ste12}
{Steiner} J.~F.,  {McClintock} J.~E.,  2012, \mn@doi [\apj]
  {10.1088/0004-637X/745/2/136}, \href
  {http://adsabs.harvard.edu/abs/2012ApJ...745..136S} {745, 136}

\bibitem[\protect\citeauthoryear{{Steiner}, {McClintock}, {Orosz}, {Remillard},
  {Bailyn}, {Kolehmainen}  \& {Straub}}{{Steiner} et~al.}{2014}]{ste14}
{Steiner} J.~F.,  {McClintock} J.~E.,  {Orosz} J.~A.,  {Remillard} R.~A.,
  {Bailyn} C.~D.,  {Kolehmainen} M.,   {Straub} O.,  2014, \mn@doi [\apjl]
  {10.1088/2041-8205/793/2/L29}, \href
  {http://adsabs.harvard.edu/abs/2014ApJ...793L..29S} {793, L29}

\bibitem[\protect\citeauthoryear{{Str{\"u}der} et~al.,}{{Str{\"u}der}
  et~al.}{2001}]{str01}
{Str{\"u}der} L.,  et~al., 2001, \mn@doi [\aap] {10.1051/0004-6361:20000066},
  \href {http://adsabs.harvard.edu/abs/2001A%26A...365L..18S} {365, L18}

\bibitem[\protect\citeauthoryear{{Svensson} \& {Zdziarski}}{{Svensson} \&
  {Zdziarski}}{1994}]{sve94}
{Svensson} R.,  {Zdziarski} A.~A.,  1994, \mn@doi [\apj] {10.1086/174934},
  \href {http://adsabs.harvard.edu/abs/1994ApJ...436..599S} {436, 599}

\bibitem[\protect\citeauthoryear{{Takahashi} et~al.,}{{Takahashi}
  et~al.}{2010}]{tak10}
{Takahashi} T.,  et~al., 2010, in Space Telescopes and Instrumentation 2010:
  Ultraviolet to Gamma Ray. p. 77320Z (\mn@eprint {arXiv} {1010.4972}),
  \mn@doi{10.1117/12.857875}

\bibitem[\protect\citeauthoryear{{Titarchuk}}{{Titarchuk}}{1994}]{tit94}
{Titarchuk} L.,  1994, \mn@doi [\apj] {10.1086/174760}, \href
  {http://adsabs.harvard.edu/abs/1994ApJ...434..570T} {434, 570}

\bibitem[\protect\citeauthoryear{{Turner} \& {Pounds}}{{Turner} \&
  {Pounds}}{1988}]{tur88}
{Turner} T.~J.,  {Pounds} K.~A.,  1988, \mnras, \href
  {http://adsabs.harvard.edu/abs/1988MNRAS.232..463T} {232, 463}

\bibitem[\protect\citeauthoryear{{Vasudevan}, {Mushotzky}, {Reynolds},
  {Fabian}, {Lohfink}, {Zoghbi}, {Gallo}  \& {Walton}}{{Vasudevan}
  et~al.}{2014}]{vas14}
{Vasudevan} R.~V.,  {Mushotzky} R.~F.,  {Reynolds} C.~S.,  {Fabian} A.~C.,
  {Lohfink} A.~M.,  {Zoghbi} A.,  {Gallo} L.~C.,   {Walton} D.,  2014, \mn@doi
  [\apj] {10.1088/0004-637X/785/1/30}, \href
  {http://adsabs.harvard.edu/abs/2014ApJ...785...30V} {785, 30}

\bibitem[\protect\citeauthoryear{{Walton}, {Nardini}, {Fabian}, {Gallo}  \&
  {Reis}}{{Walton} et~al.}{2013}]{wal13}
{Walton} D.~J.,  {Nardini} E.,  {Fabian} A.~C.,  {Gallo} L.~C.,   {Reis} R.~C.,
   2013, \mn@doi [\mnras] {10.1093/mnras/sts227}, \href
  {http://adsabs.harvard.edu/abs/2013MNRAS.428.2901W} {428, 2901}

\bibitem[\protect\citeauthoryear{{Zoghbi}, {Fabian}, {Uttley}, {Miniutti},
  {Gallo}, {Reynolds}, {Miller}  \& {Ponti}}{{Zoghbi} et~al.}{2010}]{zog10}
{Zoghbi} A.,  {Fabian} A.~C.,  {Uttley} P.,  {Miniutti} G.,  {Gallo} L.~C.,
  {Reynolds} C.~S.,  {Miller} J.~M.,   {Ponti} G.,  2010, \mn@doi [\mnras]
  {10.1111/j.1365-2966.2009.15816.x}, \href
  {http://adsabs.harvard.edu/abs/2010MNRAS.401.2419Z} {401, 2419}

\makeatother
\end{thebibliography}
%
%
%
%
\end{document}